\documentclass[aip,apl,amsmath,amssymb,reprint]{revtex4-1}
\usepackage{graphicx}
\usepackage{times}
\usepackage{color}

\usepackage{multirow}
\usepackage{makecell}

\begin{document}


\title{A highly accurate measurement of resonator $Q$-factor and resonance frequency}


\author{B. Gy\"{u}re-Garami}
\affiliation{Department of Physics, Budapest University of Technology and Economics and MTA-BME Lend\"{u}let Spintronics Research Group (PROSPIN), Po. Box 91, H-1521 Budapest, Hungary}

\author{O. S\'agi}
\affiliation{Department of Physics, Budapest University of Technology and Economics and MTA-BME Lend\"{u}let Spintronics Research Group (PROSPIN), Po. Box 91, H-1521 Budapest, Hungary}

\author{B. G. M\'{a}rkus}
\affiliation{Department of Physics, Budapest University of Technology and Economics and MTA-BME Lend\"{u}let Spintronics Research Group (PROSPIN), Po. Box 91, H-1521 Budapest, Hungary}

\author{F. Simon}
\email[Corresponding author: ]{f.simon@eik.bme.hu}
\affiliation{Department of Physics, Budapest University of Technology and Economics and MTA-BME Lend\"{u}let Spintronics Research Group (PROSPIN), Po. Box 91, H-1521 Budapest, Hungary}

\date{\today}
\begin{abstract}

The microwave cavity perturbation method is often used to determine material parameters (electric permittivity and magnetic permeability) at high frequencies and it relies on measurement of the resonator parameters. We present a method to determine the $Q$-factor and resonance frequency of microwave resonators which is conceptually simple but provides a sensitivity for these parameters which overcomes those of existing methods by an order of magnitude.
The microwave resonator is placed in a feedback resonator setup, where the output of an amplifier is connected to its own input with the resonator as a band pass filter. After reaching steady-state oscillation, the feedback circuit is disrupted by a fast microwave switch and the transient signal, which emanates from the resonator, is detected using down-conversion. The Fourier transform of the resulting time-dependent signal yields directly the resonance profile of the resonator. Albeit the method is highly accurate, this comes with a conceptual simplicity, ease of implementation and lower circuit cost. We compare existing methods for this type of measurement to explain the sensitivity of the present technique and we also make a prediction for the ultimate {\color{black}accuracy} for the resonator $Q$ and $f_0$ determination.
\end{abstract}
\maketitle

\section{Introduction}
Cavity perturbation measurements \cite{buravov71,Gruner1} are widely used to determine the electric and magnetic properties of materials at microwave frequencies. This yields the technologically important parameters including conductivity, dielectric permittvity, and magnetic permeability \cite{chen2004microwave}. The cavity perturbation technique has the clear advantage of having a higher electric or magnetic field at the sample than a non-resonant measurement, which leads to enhanced sensitivity for the material parameters. This is even more important when only small sample amounts are available. In addition, resonators allow to measure samples in a purely electric \emph{or} magnetic field, which allows to distinguish between the effects of different physical parameters. A disadvantage of the method is that results at a fixed frequency are obtained. 

Precise measurement of microwave resonator properties is also important for diverse branches of sciences and applications. Besides microwave impedance measurements \cite{Gruner1,Gruner2,Gruner3}, microwave resonators are used in particle accelerators \cite{mw_accelerator}, electron spin resonance spectroscopy and imaging \cite{PooleBook,EPRImaging}. Microwave resonators are used as filters and as essential components of microwave sources \cite{LuitenFreqStab}, radar applications \cite{pozar2004microwave}, and heating. A single-mode microwave resonator sustains a standing wave pattern at its eigen-frequency, $f_0$. The bandwith of the resonator, $\Delta f$, is characterized by its quality factor $Q$, which is the ratio of the resonance frequency and the bandwidth expressed as FWHM. The resonance curve of a {\color{black}single-mode} microwave resonator is a Lorentzian, as the time-domain transient of a resonator is an exponential function.

Most methods measure the properties of a resonator in the frequency domain \cite{LuitenReview,KajfezReview}. A straightforward method is sweeping the frequency of a source while measuring the power reflected from (the $S_{11}$ parameter) or transmitted through the resonator (the $S_{21}$ parameter). The power is detected with a detector and the obtained Lorentzian is fitted using a computer and the fitting parameters yield the eigen-frequency and the quality factor of the resonator \cite{LuitenHiResQMeas}. This method was improved with the use of vector-network analyzers \cite{KajfezVNA} which yield the reflected signal phase in addition to its {\color{black}magnitude}. The frequency swept methods have low accuracy for $Q$ and $f_0$ (Refs. \onlinecite{PetersanAnlage,LuitenReview,KajfezReview}) which could be improved with the use of a synthesized source with improved frequency accuracy \cite{LuitenHiResQMeas}. A clear drawback of the latter method is its low speed (due to the relatively long settling time of frequency synthesis) and that most of the measurement time is spent on measuring the vicinity of the resonance. {\color{black}A good example, when high speed $Q$ and $f_0$ determination is desired (down to 1 $\mu$s), is the study of light-induced photocurrent decay in semiconductors \cite{Subramanian}.}

Improved methods to measure resonator parameters use automatic frequency control (AFC) to match the source frequency to the resonator eigen-frequency \cite{Gruner2,Mehring}. This frequency is measured using a frequency counter and $Q$ values are available through the measurement of the higher harmonic components of the AFC feedback signal. Although AFC methods provide improved signal-to-noise ratio, they are rather complex and expensive to build, difficult to operate, and are prone to errors related to parasitic reflections.

An approach, which combines oscillator stability, {\color{black}reduces} measurement time, and the required instrumentation is kept at a moderate level, is that operating in the time domain \cite{Gallagher,KomachiTanaka,Amato,EatonTransient}. Rather than sweeping the frequency of a source, a resonator is probed with a pulsed carrier signal, whose frequency is close to the resonator eigen-frequency. The frequency of the transient response of the resonator matches $f_0$ (Ref. \onlinecite{SchmittZimmer}) while the time constant of the decay is related to $Q$ as: $\tau=Q/2\pi f_0$. {\color{black}The downconverted transient signal is Fourier transformed to yield the resonator parameters}. Generally speaking, measurements in the time domain have two advantages: improved accuracy (or Connes advantage \cite{Connes}) as the measurement is traced back to a stable frequency and simultaneous measurement (also known as Fellgett or multiplex advantage \cite{Fellgett}) of the whole resonator response. 

We demonstrated previously that this scheme can be successfully implemented to determine $f_0$ and $Q$ (Ref. \onlinecite{GyureRSI}) with similar accuracy than the conventional frequency domain based methods. In this respect, the time-domain measurement of the resonator parameters mimics the advance of Fourier transform NMR and IR spectroscopy which allowed for a significant revolution of these methods\cite{Ernst}. {{\color{black}We note that optical cavity ring-down spectroscopy was pioneered by Hodges and co-workers \cite{optics1,optics2,optics3,optics4} and is a well-developed analogue to the time-domain methods in microwave spectroscopy.}

However, the time-domain method also requires \textit{a priori} knowledge of the resonator resonance frequency otherwise the transient signal can be small which can affect its sensitivity for the $Q$ and $f_0$ parameters. We present an improvement of the time domain based resonator method that yields even better sensitivity for the resonator parameters. It is relatively simple to implement and it can be conceptually adapted to any radiofrequency range. Rather than {\color{black}exciting} the resonator from an external source, the resonator itself is the frequency filtering element of a feedback oscillator, it thus tunes itself to resonance. During continuous wave operation, the resonator sustains a large electromagnetic energy, which is emitted upon suddenly switching off the feedback circuit. The emitted radiation is downconverted with a stabilized oscillator and is detected with quadrature demodulation and a Fourier transformation yields the resonance curve whose parameters are readily determined. The method allows the measurement of material parameters with an unprecedented accuracy, which is explained by a detailed consideration of the different sources of noise. We also make a prediction for the ultimate sensitivity of the $Q$ and $f_0$ measurement.


\section{Measurement setup and its properties}

\begin{figure}[htp]
\begin{center}
\includegraphics[width=0.45\textwidth]{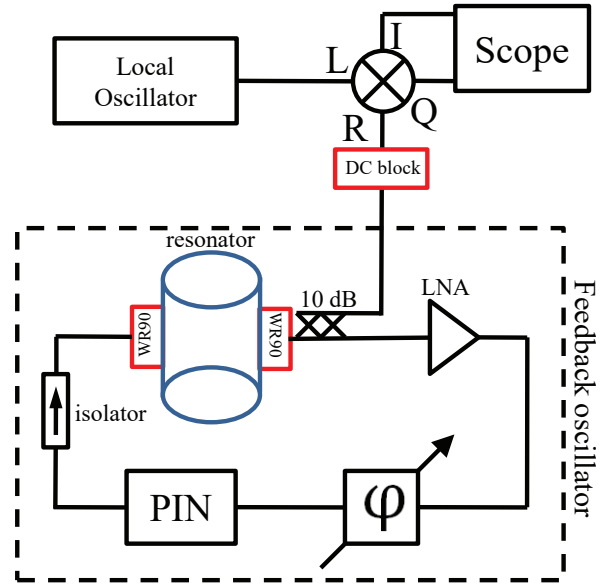}
\caption{Schematics of the experimental setup for the accurate measurement of resonator $Q$ and $f_0$, which is part of a feedback resonator. A low noise amplifier (LNA) and a phase shifter is indicated. A PIN diode stops the operation of the oscillator and the resulting transient signal is forwarded with a coupler toward the IQ mixer. The local oscillator is a PLL stabilized synthesized source. All microwave cables are coaxial SMA except for the two WR90 waveguide-coax adapters which are indicated and a DC block before the R input of the mixer.}
\label{Fig:BlockDiagram}
\end{center}
\end{figure}

Our setup for the accurate measurement of $Q$ and $f_0$ is shown in Fig. \ref{Fig:BlockDiagram}. This circuit implements a so-called feedback oscillator (FBO). FBO is a general type of stabilized oscillators which is widely used in various applications including the design of ultra-low noise oscillators incorporating dielectric \cite{nelson2004ultra} or YIG resonators \cite{odyniec1990oscillator}. The noise characteristics of FBO's are known to be well described by the Leeson's equation \cite{Leeson} that considers thermal noise related phase fluctuations.

The feedback loop consists of the transmission type resonator under test, a low-noise amplifier, a phase shifter, a PIN diode switch, and a 10 dB directional coupler (Narda 4015 C-10). We employ a cylindrical copper TE011 microwave cavity resonator with $f_0\approx 9.4\,\text{GHz}$ and an unloaded $Q_0\approx 10,000$. This resonator can be used to measure microwave properties of materials with the cavity perturbation technique. The resonator is connected to WR90 waveguide-coax adapters with two coupling irises as the circuit consists of coaxial parts. The resonator is undercoupled for both the input and output with $S$ parameters of $S_{11}=S_{22}= 3\,\text{dB}$ which results in a transmission loss $S_{21}= 6\,\text{dB}$. These coupling parameters represent a compromise between the transmission loss and the sensitivity of the $Q$ measurement to material parameters \cite{buravov71,Gruner1} as we describe in the Supplementary Material.

The LNA is custom made (Janilab Inc.) with NF=1.4 dB, Gain=15 dB and 1 dB compression point, P1dB=10 dBm. Microwaves are switched with a reflection type fast PIN diode with a less than 5 ns 10-90\% rise-fall transient time (Advanced Technical Materials, S1517D). The PIN diode is driven by an arbitrary waveform generator (HP33120A) with varying pulse length and frequency. The transient signal of the feedback loop is measured through a directional coupler whose output is connected to either a power detector (HP8472A) or to the RF port of an IQ mixer (Marki IQ0618LXP) for downconversion. {\color{black}We took care to operate the mixer in its linear domain, i.e. to avoid its saturation due to high power input.} It is important to electrically isolate the mixer from the rest of the circuit: to this end, we employ two facing WR90 waveguide-coaxial adapters, which are separated by a Mylar foil and are connected with plastic screws. This DC blocking element is placed before the R input of the mixer. {\color{black}In addition, a microwave isolator before the resonator is required to dissipate the power reflected from the switched off PIN diode}.

The local oscillator is a PLL stabilized synthesized frequency source (Agilent HP83751B, 2-20 GHz or Kuhne Electronic MKU LO 8-13 PLL), which provide the LO power of 10 dBm. A high-frequency and high-end oscilloscope with an OCXO stabilized clock (bandwidth 1 GHz, Rohde \& Schwarz RTO1014) is required to measure the fast ($\sim 100$ ns) transients and it is capable of averaging several transients with a high repetition rate (typically 100 kHz) with essentially zero dead time, i.e. without the loss of triggers. The switch on mechanism of the FBO is somewhat involved as it depends on the loop gain, phase etc. It is discussed in the Supplementary Material. In turn, the switch off transient signal solely depends on the resonator ringdown, therefore we focus on it in the following.

The conditions for the feedback oscillator operation are that i) the LNA gain overcomes the losses in the circuit (including the transmission loss in the resonator) and ii) the phase of the returning microwave signal matches that in the resonator. These are also commonly known as the Barkhausen's criteria. The stable FBO operation occurs due to the LNA saturation (also known as gain compression) as it reduces the net LNA gain and the stable operation occurs when the compressed LNA gain equals exactly the losses. Upon compression, third harmonic production occurs on the LNA output, however some of the elements (e.g. the waveguide to coax adapters) in our circuit effectively suppress it.

The phase matching condition is enabled by the mechanical phase shifter (Arra AR4329) and a high stability FBO requires an automatic control of the phase match \cite{NelsonNISTConf}. The transmission cavity also acts as a phase shifter and the magnitude of phase shift depends on the frequency difference between the FBO frequency and the cavity eigen-frequency. Therefore the FBO operates even when the phase matching is imperfect, however at the cost of detuning the FBO frequency from the resonator eigen-frequency, which results in a less stable operation and a lower sustained power in the circuit. The latter effects are the result of an increased resonator transmission loss when the FBO frequency does not match the resonator eigen-frequency. Then, the LNA compression decreases (and the gain increases) in order to compensate for the larger transmission loss, which results in a decreased net power in the circuit. The influence of the phase mismatch on the FBO performance is important for the later considerations.

When the net amplification in the loop is greater than 0 dB, the small amount of microwave radiation in the loop is amplified and the transmission through the resonator filters the frequency. As the power in the loop reaches saturation of the amplifier, the net amplification decreases until it reaches 0 dB and a stationary state is reached. 

\begin{figure}[htp]
\begin{center}
\includegraphics[width=0.45\textwidth]{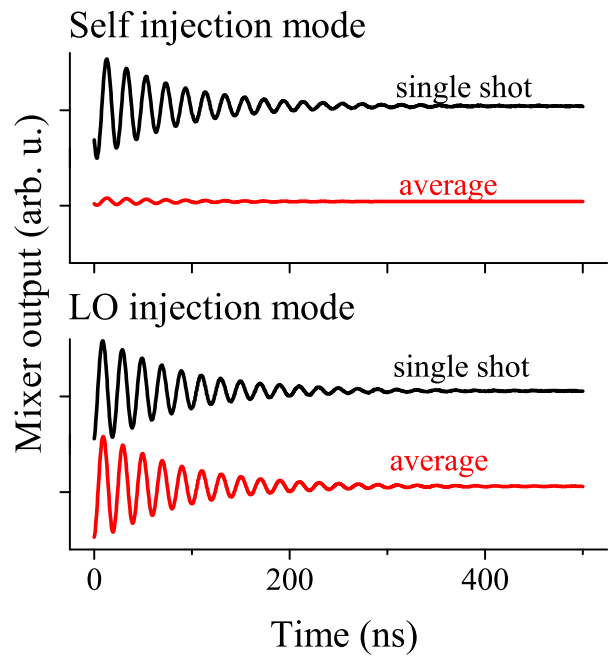}
\caption{Single shot and averaged cavity transients when the feedback oscillator is either in the LO injection locked mode or self-injected from the previous cycle. Note that for self-injection, the FBO transients average to zero, as phases in each transient are uncorrelated with respect to the local oscillator. However in the LO injection locked mode, the phase between the FBO and the local oscillator is constant.}
\label{Fig:2states}
\end{center}
\end{figure}

When a properly set up feedback loop is switched on, the loop may be in one of three distinct states. 1) Noise injection; this mode of operation starts from the thermal noise of the LNA output, that is filtered by the resonator and is amplified in each consecutive amplifying rounds. The LNA output has a power spectrum of -174 dBm/Hz+Gain+NF$\approx -157$ dBm/Hz. This operation occurs when no LO is present and the FBO output is monitored with a power detector. The FBO startup is inherently random due to the random nature of the noise power on the resonator frequency that is demonstrated in the Supplementary Material. 2) Self-injection; if not sufficiently long time elapses after the PIN diode switch-off before the next switch-on, the residual power emitted from the resonator injects it again. The emitted power decays as $\text{e}^{-t/\tau}$, where $\tau=Q/\omega_0$ (e.g. $\tau=160\,\text{ns}$ for $Q=10,000$ and $f_0=10\,\text{GHz}$) and that this mode of operation is realized if the switch-off time is not longer than $2-3\,\mu\text{s}$ in our case. 3) LO injection; we cannot fully eliminate the radiation leakage from the LO towards the circuit whose power can be larger than the residual power during the circuit switch-off. This leads to an LO injected operation of the FBO, i.e. the FBO starts operating in phase with the LO after the next switch-on. This occurs when the switch-off time is longer than the above mentioned $2-3\,\mu\text{s}$ thus the self-injection is no longer active. We chose a 10 $\mu\text{s}$ delay between the pulses, in order to operate the FBO in this mode for the reasons detailed below. We estimate that the LO leakage cannot be reduced below about 50-90 dB in our experimental conditions.

The effects of the LO injection and the self-injection modes on the FBO transients is demonstrated in Fig. \ref{Fig:2states}. We chose an IF of 50 MHz, for both type of operations, which leads to the oscillatory nature of the transient. We found that the phase of the downconverted consecutive FBO transients is changing for the self-injection operation. This is the result of the FBO phase being arbitrary as compared to the LO. As a result, the FBO transients average to zero. However for the LO injection mode, the consecutive pulses have the \emph{same} phase after downconversion with respect to the LO. In fact, the LO injection mode is advantageous for the intended operation as it enables signal averaging, thus it allows to reduce the signal noise. We found that an IF of $20-50\,\text{MHz}$ is a good choice as it lies above the "knee-point", where the output noise of the mixer is minimal. The oscilloscope and mixer IF bandwidths prevent the use of a much larger IF frequency. 

We note at this point that the use of FBO differs conceptually from our previous time domain method in Ref. \onlinecite{GyureRSI}: the present method acts as if two independent frequency sources were present with adjustable frequencies that allows the use of a high IF. In the previous work, the same LO was used to irradiate and downconvert the transient signal. This means that either the transient signal is large and the IF is low (when the LO is near the resonator $f_0$), which gives rise to a noisy signal. Another choice is a small transient signal and a large IF (when the LO is offset from the resonator $f_0$ by about $10-100$ MHz). Clearly, a compromise is to be met for both cases, which could in principle be avoided with the use of two independent (but preferably phase locked) LO sources.

\begin{figure}[htp]
\begin{center}
\includegraphics[width=0.45\textwidth]{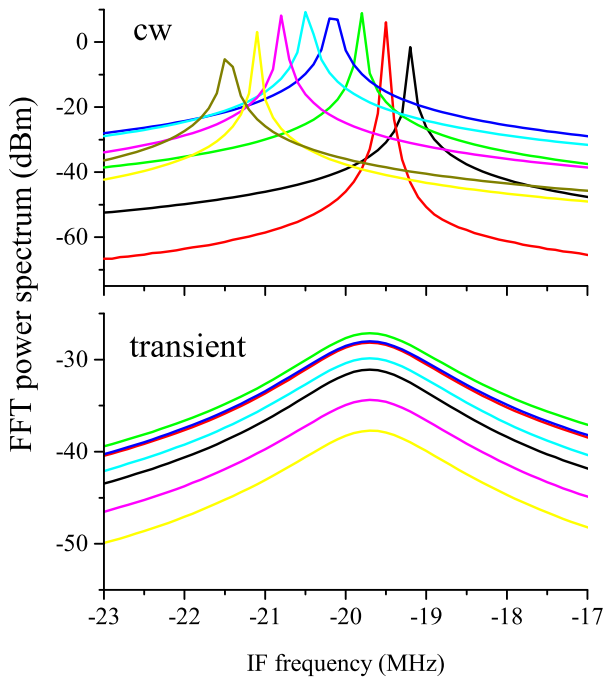}
\caption{FFT power spectra of the FBO for continuous wave operation (upper panel) and for the transients (lower panel) while the return phase of the FBO is varied. Curves with the same color correspond to the same return phase. Note that while the cw frequency changes considerably, the transient frequency remains constant, while its power changes. Note that the cw FBO signal is resolution limited and that the large width of the transient spectra is due to the quality factor of the microwave resonator. The scale for the transient corresponds to the temporary power and its time average would be inevitably smaller.}
\label{Fig:phase_plot}
\end{center}
\end{figure}

We studied the output of the FBO in continuous wave (cw) and transient operation. For the earlier, the FBO is freely running and the downconverted signal is digitized with the oscilloscope and Fourier transformed. The power spectrum of the Fourier transform is equivalent to a heterodyne spectrum analysis of the FBO output that was verified with a HP8566B spectrum analyzer. The result is shown in the upper panel of Fig. \ref{Fig:phase_plot}; the oscillator output is limited by the digitization (or FT) resolution and the maximum output power of the FBO is about 10 dBm which matches well with the 1 dB compression point of the LNA (P1dB=10 dBm). To study the FBO transients, we operated the circuit in the LO injection mode. The transient signal is triggered with the PIN diode switch off signal. 

We mentioned that the frequency of the FBO depends sensitively on the return phase in the circuit when it is operated in a continuous wave (cw) operation. This is demonstrated in the upper panel of Fig. \ref{Fig:phase_plot}: we employ a resonator with a FWHM of $\sim 1\,\text{MHz}$ and the FBO frequency can be detuned with respect to the resonator eigen-frequency by as much as 2 MHz by altering the return phase. This results in a reduced power in the FBO and in a somewhat increased phase noise of the oscillator. This problem can be overcome by an electronic servo control of the return phase, which results in a highly stable FBO operation as described in Ref. \onlinecite{NelsonNISTConf}. However, this sensitivity of the frequency on the return phase prevents any meaningful measurements of the resonator eigen-frequency when it is used as part of a (cw) FBO. 

Our approach to study the FBO transients solves this problem: when the FBO is switched off as described above, the energy accumulated in the resonator is emitted on the resonator eigen-frequency \cite{GyureRSI}. This is clearly demonstrated in the lower panel of Fig. \ref{Fig:phase_plot}: the signal emitted during transients remains at the \emph{same} frequency even for a significant return phase mismatch, which strongly detunes the FBO frequency in the cw operation. The constant carrier frequency of the transient can thus be identified as the resonator eigen-frequency.

Besides being immune to detuning effects due to the return phase mismatch, the transient mode operation has additional advantages concerning the measurement noise: the FBO noise depends strongly on the amplifier noise figure, which is often inferior in the presence of a large compression, an effect that is described by Leeson's equation \cite{Leeson}. However, the LNA does not affect the FBO transient signal, therefore it can be regarded as a perfect oscillator output besides the transient nature of the signal. 

\section{The error of the resonator measurements}

In the following, we discuss the practical use of the FBO transients for the measurement of $Q$ and $f_0$ and also evaluate the measurement performance. 
\begin{table*}[htp]
\begin{center}
    \begin{tabular*}{0.8\textwidth}{@{\extracolsep{\fill}}lllll}
    \hline \hline
		Method&\makecell[l]{frequency\\accuracy}&IF&detection method&Refs.\\
		\hline
		\multirow[l]{2}{*}{Frequency sweep}&\multirow[l]{2}{*}{low}&\multirow[l]{2}{*}{$\sim10$ kHz}&power detector&\onlinecite{PetersanAnlage,LuitenReview,KajfezReview}\\
		&&&mixer (VNA)&\onlinecite{LeongVNA1,LeongVNA2}\\
		AFC based&medium&$\sim10$ kHz&power detector&\onlinecite{Gruner2,HolczerPRB,Mehring}\\
		Stepped frequency sweep&high& $< 100$ Hz&power detector&\onlinecite{LuitenHiResQMeas}\\
		Time domain&high&\makecell[l]{$< 1$ MHz\\$20-50$ MHz}&\makecell[l]{power detector\\mixer}&\makecell[l]{\onlinecite{Gallagher,Amato,EatonTransient}\\\textbf{present work}}\\
		\hline \hline
    \end{tabular*}
    \caption{The methods used for the measurement of resonator $Q$ and $f_0$ and their most important characteristics including frequency accuracy, intermediate frequency value, and detection method.}
    \label{tab:MethodsTable}
\end{center}
\end{table*}
We believe that the resonator measurement methods can be classified into 4 main groups as shown in Table \ref{tab:MethodsTable}: i) frequency swept methods \cite{PetersanAnlage,LuitenReview,KajfezReview}, ii) automatic frequency control (AFC) based methods \cite{Gruner2,HolczerPRB,Mehring}, iii) PLL stabilized frequency stepped methods \cite{LuitenHiResQMeas}, and iv) time domain methods \cite{Gallagher,KomachiTanaka,Amato,EatonTransient,GyureRSI} such as the present work.

In frequency swept methods (i) the reproducibility and accuracy of the frequency is inherently low. We can estimate the error of such measurements (presented in the Supplementary Material) from the frequency inaccuracy which is provided by the sweeper oscillator manufacturers. The AFC based methods (ii) keep the frequency of an oscillator on the resonator $f_0$, which allows a direct frequency counting of $f_0$ but $Q$ is accessible only indirectly and after calibrating measurements \cite{Gruner2,HolczerPRB,Mehring}. Although frequency accuracy of the AFC is better than that of the frequency swept methods, its stability is still inferior compared to the PLL performance. Most reports using methods i-ii employ power detectors that are known to have a larger noise and lower dynamic range than mixers \cite{PooleBook,pozar2004microwave}. The use of vector network analyzers (i.e. phase sensitive mixing) lead to improved accuracy of frequency swept methods \cite{LeongVNA1,LeongVNA2}. The PLL based stepped frequency method \cite{LuitenHiResQMeas} represents an improvement in frequency stability, however the relatively long PLL stabilization time (typically $10\,\text{ms}/\text{point}$) limits the attainable accuracy. 

Another important factor is the magnitude of the intermediate frequency. Detectors and mixers have a non-constant noise behavior which is often approximated by a $1/f$ characteristics (also known as flicker noise), where $f$ is the frequency difference between the carrier and the studied sideband frequency. The $1/f$ behavior is followed by a constant noise value (or thermal noise floor) above the so-called "knee-point". As a result, either detectors or mixers are best operated at the highest possible intermediate frequency. The bandwidth of power detectors is usually limited to a few MHz but our mixers have a 500 MHz IF bandwidth and it allows to choose an appropriately high IF. The time domain methods, that we are aware of \cite{Gallagher,Amato,EatonTransient}, employed power detectors, which limited the measurement for rapid resonator transients. The present time domain method combines the frequency accuracy of a PLL system with the low noise of mixers for a high value of the IF.

We previously set novel benchmarks which enabled to compare different methods of the resonator $Q$ and $f_0$ measurement\cite{GyureRSI} even for resonators with orders of magnitude $Q$ variation. We found that the error of the $Q$ and $f_0$ measurement can be defined as:

\begin{align}
\delta\left(Q\right) := \frac{\sigma \left(Q\right)}{\overline{Q}}\,; \delta\left(f_0\right):=\frac{\sigma \left(f_0\right)}{\overline{\Delta f}},
\label{Eq:ErrorDefinition}
\end{align}
where $\overline{Q}$ and $\overline{\Delta f}$ are the mean values of $Q$ and the resonator bandwidth $\Delta f$, respectively. When comparing different measurement methods, a normalization with the measurement time is also important and we present data which is normalized to 1 seconds. 

\begin{table}[htp]
\begin{center}
    \begin{tabular*}{0.45\textwidth}{@{\extracolsep{\fill}}lllll}
    \hline \hline
    Method & $Q$ & $t\,\text{[s]}$ & $\delta\left(Q\right)$ & $\delta\left(f_0\right)$\\ \hline
    Ref. \onlinecite{LuitenHiResQMeas} & $10^{8}-10^{9}$ & 10 & $6\times 10^{-4}$ & $6\times 10^{-4}$ \\
    Ref. \onlinecite{Mehring} & $2.5\times 10^{4}$ & 3 & $10^{-3}$ & $10^{-3}$\\
    Our previous method & $10^4-10^5$ & 1 & $10^{-3}$ & $10^{-3}$\\
		Present method & $10^4$ & 1 & $\mathbf{6\times 10^{-5}}$ & $\mathbf{6\times 10^{-5}}$\\
    \hline \hline
    \end{tabular*}
    \caption{Comparison of the error of the different $Q$ and $f_0$ measurement methods for various $Q$ values. The measurement duration, $t$ is also given in seconds.}
    \label{tab:Comparison}
\end{center}
\end{table}

We present a comparison for these error definitions in Table \ref{tab:Comparison} for various methods. It is clear from the table that these are useful error definitions as these provide a $Q$ independent measure and we also note that $\delta\left(Q\right)\approx \delta\left(f_0\right)$ for \emph{all} methods. This latter property can be proven considering that $Q=f_0/\Delta f$ and employing conventional error propagation as follows:
\begin{gather}
\delta\left(Q\right) = \frac{\sigma \left(Q\right)}{\overline{Q}}=\frac{\sigma \left(f_0\right)}{\overline{f_0}}+\frac{\sigma \left(\Delta f\right)}{\overline{\Delta f}}.
\end{gather}
Here, the first term can be neglected as $\overline{f_0}\gg \sigma \left(f_0\right)$ and we show in the Supplementary Material that $\sigma \left(\Delta f\right)\approx \sigma \left(f_0\right)$.

We measure the signal with a 100 kHz repetition time with essentially no dead-time, which allows to accumulate it several times before read out. We varied the measurement time between 100 ms and 1 sec, the acquired signals are read-out, Fourier transformed and Lorentzian fits to the data yield $Q$ and $f_0$. By measurement time, we mean the total time spent on acquiring the signal, transferring it to computer and analyzing the data until the relevant resonator parameters are stored. The effective time spent on measuring the signal is about only 5-10\% of the measurement time. Standard deviations, $\sigma \left(Q\right)$ and $\sigma \left(f_0\right)$, of the relevant data are determined from the conventional definitions and the result for the present method is also given in Table \ref{tab:Comparison}.

Table \ref{tab:Comparison} shows that all methods, including our previous transient work \cite{GyureRSI}, provide a $\delta\left(Q\right)\approx\delta\left(f_0\right)\approx 10^{-3}$. In contrast, when normalized to a 1 second measurement time, the present FBO transient based method yields an order of magnitude improvement of $\delta\left(Q\right)=\delta\left(f_0\right)\approx 6\times 10^{-5}$. We believe that the significant improvement presented by our method is due to the use of time domain, a stable LO, and the optimal use of averaging.

We performed Monte Carlo simulations (details are given in the Supplementary Material) in order to explain the improved error and also to understand the limits of the resonator parameter measurements. We generated transient signals which mimic the behavior of the measurements with added signal noise, and frequency or phase noise of the LO. We assumed a Gaussian signal noise with a given standard deviation, $\sigma_{\text{s}}$, superimposed on the decaying transient. Concerning the LO frequency and phase noise, we also assumed that these parameters have a Gaussian distribution around a mean value, which may be oversimplifying given that oscillator noise has usually more complicated characteristics \cite{Leeson,Daud,RutmanPhaseNoiseAnalysis}. We found that the LO frequency and phase noise affects differently the error of the $f_0$ and $Q$, which is not supported by the experimental observation. However, the added signal noise results in the same magnitude for the two types of errors, which is consistent with the observed data, we therefore conclude that the error originates from a signal noise of the detected voltage.

The Monte Carlo simulation indicates that the observed errors of $Q$ and $f_0$ are obtained when the initial peak-to-peak value of the transient is $\sim 10^{4}$ larger than $\sigma_{\text{s}}$, which is referred to as signal-to-noise ratio (SNR) in the following. A careful inspection of our raw data confirms that we do observe a signal with this SNR. As a next step, we identified that the oscilloscope digital noise is the origin of this: our oscilloscope has a digit noise of $\sim 1:500$ as compared to the full span (for signals above $50\,\text{mV}_{\text{pp}}$ which is the case herein). Given that the span cannot be optimal for an arbitrary signal, we estimate a $\sim 1:200$ digital noise as compared to the transient signal peak-to-peak value. For a digital averaging with about 1,000-10,000 scans, this is reduced to $\sim 1:10^{4}$, which matches well the experimentally observed signal-to-noise ratio. We note that we cannot average for more than the above values during a 1 second measurement time given that the transients are measured for about $10-50\,\mu\text{s}$ and the majority of the time is spent with the data transfer, signal analysis and data storage. If we considered the shorter effective measurement time, we would artificially obtain and even smaller error of the resonator parameter determination. 

Our setup has three shortcomings which could be improved in the future: i) the large, 1 GHz bandwidth, which cannot be arbitrarily set for the oscilloscope, ii) the presence of the digital noise, and iii) that a general purpose oscilloscope has is noisier input than a low-noise amplifier even for small signal inputs. The oscilloscope has an RMS noise of $100\,\mu\text{V}$ for small signals, which corresponds to a noise figure (NF) of $11\,\text{dB}$. A high resolution digitizer card with ample oversampling for bandwidth reduction and control, combined with a low noise IF amplifier, could improve the measurement. The optimal equivalent noise bandwidth (ENBW), which directly affects the SNR, can be obtained from considering that the acquisition frequency is dictated either by the working IF or the resonator bandwidth, $\Delta f$. 

Assume that the dominant factor is $\Delta f$, then a sampling frequency of about $10\times \Delta f$ is needed for at least a 100 data points, which takes $t=10/\Delta f$. This yields that during a 1 second averaging, $\Delta f/10$ scans can be acquired, which leads to an ENBW=100 Hz, which is independent of $\Delta f$. We believe that a more realistic value is ENBW=1 kHz due to the time used for data transfer, fitting, etc. This consideration breaks down for cases when the resonator bandwidth is low (below about 10 kHz) and a large IF (above 1 MHz) is to be used due to the mixer "knee-point". 

As a result, under ideal circumstances (true thermal noise, optimized ENBW, a nearly ideal IF amplifier with NF $\sim 1\,\text{dB}$), the noise could be as low as 30 nV (i.e. the thermal floor of -174 dBm/Hz+31 dB, of which 30 dB due to the 1 kHz ENBW). We envisage a transient signal with a 0 dBm starting amplitude measured with typical mixer with conversion loss of 10 dB. This results in a $-10\,\text{dBm}$ or $200\,\text{mV}_{\text{pp}}$ signal, which together with the 30 nV noise would give $\delta \left(Q\right)=\delta\left(f_0\right)\approx 10^{-7}$ that would represent 3 orders of magnitude improvement compared to our present result. 

We make two additional remarks related to the above noise considerations. First, the Schottky formula for the current shot-noise yields that its voltage noise contribution would be equal to that of the thermal (or Nyquist) noise at a -10 dBm input signal level. Second, we believe that the above minimum error of the $Q$ and $f_0$ measurement may not be reached, as at very low signal noise levels other noise sources (e.g. the LO frequency or phase noise) would dominate it. The magnitude of the latter contributions is however yet impossible to estimate.

{\color{black}We would like to emphasize that above we considered stochastic noise sources only (also known as Type A Elemental Uncertainty\cite{UncertaintyBook}). This means that our method overperforms alternatives when the noise is dominated by such sources and other factors, e.g. the repeatability of the measurements, determination of coupling factors, etc. do not play a significant role. Typical situations for this are when tiny relative variations in e.g. the sample absorption as a function of temperature or magnetic field is studied which does not involve removing and replacing the sample.}

{\color{black}Under such circumstances, an} improved accuracy in the measurement of $Q$ and $f_0$ directly translates to a larger sensitivity to material parameters, electric permittivity or magnetic permeability for the same setup, geometry of volume filling factor as we discuss it in the Supplementary Material. This in turn enables measurement on systems which involve a smaller change in these parameters, or on much smaller amounts of materials for the same resonator system. This may eventually lead to study problems which have been so far inaccessible by the conventional methods.

\section{Summary}
In summary, we presented a feedback oscillator based measurement of resonator quality factor and resonance frequency. Knowledge of these parameters is crucial when material properties are studied using radiofrequency or microwave resonators. The method is based on the detection of resonator transients clocked with a PLL stabilized LO and it yields highly stable and reproducible results for $Q$ and $f_0$. The method yields about an order of magnitude more accurate $Q$ and $f_0$ values than alternative methods. We critically compare the different measurement methods, which allows us to identify the reasons behind the enhanced accuracy, the limitations of the method, and the origins of the noise. We predict that the $Q$ and $f_0$ measurement could be further improved and we identified the necessary requirements.

\section*{Acknowledgements}
Work supported by the Hungarian National Research, Development and Innovation Office (NKFIH) Grant Nrs. 2017-1.2.1-NKP-2017-00001 and K119442.

\appendix
\newpage
\pagebreak
\clearpage

\section{The noise injected operation of the FBO}

As mentioned in the main text, the FBO can be also operated in the so-called noise injected mode. To achieve this, the mixer and the local oscillator is replaced with a power detector. Then, the FBO operation start from the amplified thermal noise on the LNA output. Given the random nature of the thermal noise, it is expected that the FBO starts in a random fashion.

\begin{figure}[htp]
\begin{center}
\includegraphics[width=0.5\textwidth]{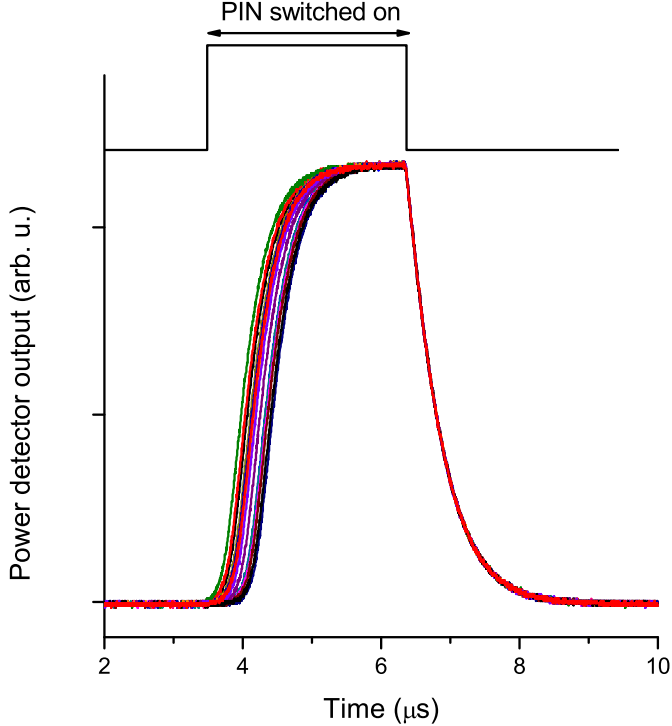}
\caption{Individual switch on and off traces for the noise injected FBO operation. Note that the switch on transients do not lay on one another, but the switch off traces are identical. The state of the PIN diode is indicated.}
\label{Fig:uncertainstart}
\end{center}
\end{figure}

This is indeed the case, as we show in Fig. \ref{Fig:uncertainstart}.: individual switch on traces appear in a random manner. In contrast, the switch off transients are always identical. Note that the switch on transient is not an exponential. The shape of this function is determined by a combination of two factors, the time constant of the cavity and the gradual saturation of the amplifier in the feedback circuit. As this function is much more difficult to interpret than the one at the switch off transient, the latter is used in the analysis.

Also note that the start of the switch on transients are uncertain. As stated before, if no other radiation is present in the loop the thermal noise (-174 dBm/Hz) is amplified. Since in this state the amplifier is not saturated, the net amplification may be greater than 0 dB for several frequencies and phases of microwave radiation. This causes these modes to 'race' and after the saturation of the amplifier only one of these modes remains. This can be visualized when one the modes has a shorter time constant than the dominant mode. The race of these modes is determined by the random nature of the thermal noise at the switch on signal.

\section{The optimal $S$ parameters for a transmission cavity}

The voltage of the transient signal for a transmission microwave cavity reads \cite{PooleBook,GyureRSI}:
\begin{equation}
V\left(t\right)  =\frac{2\sqrt{\beta_{1}\beta_{2}}}{1+\beta_{1}+\beta_{2}}  \sqrt{P_{0}Z_{0}} \; e^{-\left(1+\beta_{1}+\beta_{2}\right)\frac{\omega_{0}t}{2Q_{0}}} \label{transmission}
\end{equation}
where $\beta_{1,2}$ are the input and output coupling coefficients, respectively. The loaded quality factor, $Q$, reads for this case:
\begin{equation}
 Q=\frac{Q_{0}}{1+\beta_{1}+\beta_{2}} 
\label{Q_value}
\end{equation}
The amount of transmitted versus the incoming power is denoted with $T$. It is also known as the $S_{12}$ parameter and its magnitude reads:

\begin{equation}
T=\frac{4 \beta_{1} \beta_{2}}{\left(1+\beta_{1}+\beta_{2}\right)^{2}} 
\label{T_value}
\end{equation}
It is clear that increased coupling leads to a reduced $Q$ factor, which leads to a shorter transient and that the transmitted power is a nontrivial function of the $\beta$'s. Given the transient nature of the studied signal, one needs to optimize for the transmitted \emph{energy} during a single transient.

The condition for the maximum transmitted energy during the transient is obtained from the Fourier transform of Eq. \eqref{transmission} and its integral over the whole frequency range as follows: 
 \begin{gather}
\int_{-\infty}^{+\infty} |\tilde{V}\left(\omega\right)|^{2} \mathrm{d} \omega = \int_{- \infty}^{+ \infty} \frac{1}{2 \pi} \frac{T P_ {0} Z_{0}}{\left(\frac{\omega_{0}}{2Q}\right)^{2}+ \omega^{2}} \mathrm{d} \omega \\ = \frac{TP_{0} Z_{0} \cdot Q}{\omega_{0}}
\end{gather}
The result shows that the maximum transmitted energy per pulse occurs when $Q \cdot T$ is maximal.  A similar calculation shows that when one optimizes the experiment for the maximum transmitted microwave voltage signal, the area under $\mathsf{Re} \; \tilde{V}\left(\omega\right)$ needs to be maximized, which occurs for the maximum of $Q \cdot \sqrt{T}$.

 \begin{table}[!htbp]
 \begin{center}
 \begin{tabular}{|c|c|c|c|}
 \hline
 Measured quantity & $\beta_{\textrm{1,2}}$ & Transmission & $Q$ \\
 \hline
Power& $1$ & $4/9 \; \left(3.5 \; \textrm{dB}\right)$ & $Q_{0}/3$ \\
 \hline
Voltage  & $1/2$ &$1/4 \; \left( 6 \; \textrm{dB} \right)$ & $Q_{0}/2$ \\
 \hline
 \end{tabular}
 \end{center}
 \caption{The optimum coupling coefficients for maximizing the transmitted power or microwave voltage. The calculated transmission factor and quality factor values are also given.}
\label{Beta_table} 
\end{table}
 
The conditions for the coupling coefficients for a maximal $Q \cdot T$ or $Q \cdot \sqrt{T}$ is readily obtained from Eqs. \eqref{Q_value} and \eqref{T_value} and the result is summarized in Table \ref{Beta_table}. We give in the main text that we use the $\beta_1=\beta_2=1/2$ condition as it optimizes for the maximal microwave voltage, which is studied using a mixer.

\section{Origin of the error of the resonator frequency and quality factor measurement in frequency swept experiments}

We defined previously the following quantities to characterize the goodness of $f_0$ and $Q$ factor measurements as:
\begin{align}
\delta\left(Q\right) := \frac{\sigma \left(Q\right)}{\overline{Q}}\,, \delta\left(f_0\right):=\frac{\sigma \left(f_0\right)}{\overline{\Delta f}},
\label{Eq:SM_ErrorDefinition}
\end{align}
where $\overline{Q}$ and $\overline{\Delta f}$ are the mean values of $Q$ and the resonator bandwidth $\Delta f$, respectively. The $Q=f_0/\Delta f$ definition and standard error propagation formula yields:

\begin{gather}
\delta\left(Q\right) = \frac{\sigma \left(Q\right)}{\overline{Q}}=\frac{\sigma \left(f_0\right)}{\overline{f_0}}+\frac{\sigma \left(\Delta f\right)}{\overline{\Delta f}}.
\end{gather}
Here, the first term can be neglected as $f_0\gg \sigma \left(f_0\right)$. In what follows, we prove that $\sigma \left(\Delta f\right)\approx \sigma \left(f_0\right)$.

We consider the frequency swept methods \cite{LuitenReview,KajfezReview} as being the most common to determine these quantities. After measuring a resonator response, a Lorentzian curve:
\begin{align}
\text{Lor}_{f_0,\Delta f,A}\left(f\right)= \frac{A \Delta f}{2 \pi}\frac{1}{\left(f-f_0 \right)^2+\left(\frac{\Delta f}{2}\right)^2}.
\label{Eq:SM_Lor}
\end{align}
This function has an integrated area of $A$, a FWHM of $\Delta f$, and is centered at $f_0$. We assume that the cavity response (or non-dB S parameters) is represented as ($y_i,\,x_i$) data points, where $x_i$ denote the frequency. It is common to fit the Lorentzian to the data with the least-squares methods, which find the minimum of 

\begin{gather}
\chi^2=\sum_i\left[y_i- \text{Lor}_{f_0,\Delta f,A}\left(x_i\right)\right]^2.
\end{gather}

The standard result for the minimum of $\chi^2=\sigma^2 \cdot (n-m)$, where $n$ is the number of the data points and $m$ is the number of the fitted parameters. The standard deviation for a fitted parameter $a$ (it is either of $f_0$, $\Delta f$, or $A$) reads:

\begin{gather}
\sigma^2(a)=\sigma^2\cdot \sum_i \left( \frac{\text{d}\text{Lor}_{f_0,\Delta f,A}\left(x_i\right)}{\text{d}a}  \right)^{-2}.
\end{gather}
The latter sum can be well approximated with the improper integrals of the corresponding derivatives of the Lorentzian to $\pm \infty$ as:

\begin{gather}
\sum_i \left( \frac{\text{d}\text{Lor}_{f_0,\Delta f,A}\left(x_i\right)}{\text{d}f_0}  \right)^{2}\approx \frac{1}{\Delta x}\frac{2A^2}{\pi\Delta f^3}\\
\sum_i \left( \frac{\text{d}\text{Lor}_{f_0,\Delta f,A}\left(x_i\right)}{\text{d}\Delta f}  \right)^{2}\approx \frac{1}{\Delta x}\frac{A^2}{2\pi\Delta f^3}
\label{eq:SM_approximation}
\end{gather}
where $\Delta x$ is the interval length (supposedly uniform) between two consecutive frequency points. The approximation is good when the summation goes for a region larger than $\Delta f$, which is always satisfied in practice. The above equations show that the error of $f_0$ is always a factor two smaller than that of $\Delta f$, which is essentially the result of the $Q$ factor definition. It is recognized that the quantity $f_0$ is related to $1/2Q$ rather than $1/Q$.

Eq. \eqref{eq:SM_approximation} allows to quantitatively estimate the error of the respective parameters:
\begin{gather}
\delta\left(f_0\right)=\frac{\sigma\left(f_0\right)}{\overline{\Delta f}}=\sqrt{\frac{\pi}{2}}\frac{\sigma}{A}\sqrt{\overline{\Delta f}\Delta x}\\
\delta\left(Q\right)\approx\frac{\sigma\left(\Delta f\right)}{\overline{\Delta f}}=\sqrt{2\pi}\frac{\sigma}{A}\sqrt{\overline{\Delta f}\Delta x}
\label{SM:parameter_errors}
\end{gather}
where we substituted the respective expectation values into Eq. \eqref{eq:SM_approximation}. This result is remarkable: as we show it allows to estimate the expected error of $Q$ and $f_0$ in a parameter free way for a frequency swept experiment. 

We recognize that the last term in the above equations can be rewritten as $\sqrt{\overline{\Delta f}\Delta x}=\overline{\Delta f}/\sqrt{N}$, where $N$ is the number of measurement points per the resonance width. We believe that in a typical frequency swept experiment, one chooses a reasonable sweep width of about 10 times larger than $\overline{\Delta f}$ and a typical choice of $N$ is always the same, about $N=100\dots1000$. 
We can therefore perform a Monte Carlo simulation of the $Q$ and $f_0$ determination on a Lorentzian as follows: the experimental data is simulated by assuming that the resonator response forms a Lorentzian curve with magnitude $A$, a fixed width of $\Delta f$ and whose $f_0$ is a random variable at each $x_i$ data point (we denote it as frequency noise) and that it also contains an added signal noise, which is also a random variable. The data points generated this way are fitted with the least squares method. 
We consider the effects of the two types of noises separately.

The effect of the amplitude noise is a straightforward consideration: the amplitude (or height) of the Lorentzian scales with $A/\Delta f$, and the error of the parameters in Eq. \eqref{SM:parameter_errors} scales with its inverse. It means that the amplitude noise, which induces a $\delta\left(f_0\right)\approx 10^{-3}$, can be determined in units of the amplitude of the Lorentzian and we obtain that a signal noise whose standard deviation is about $3\cdot 10^{-3}$ of the lorentzian amplitude leads to the observed error of $f_0$ and $Q$. This amount of signal noise is reasonable in our opinion for the most measurement techniques \cite{LuitenReview,KajfezReview}.

The other potential source of the noise is the inaccuracy of the frequency during a frequency sweep (the frequency noise). While the noise inaccuracy is most probably a time varying deviation (not a scattering) from the ideal linearly swept frequency, we model it with a random variable. We find that this case, the effect of $A$ cancels from the problem and the magnitude of the frequency noise can be expressed in terms of the Lorentzian width, This means that the obtained error of $f_0$ and $Q$ remains constant if the standard deviation of the frequency noise per the lorentzian FWHM remains constant. Namely, the $10^{-3}$ error of $f_0$ can be reproduced if we assume that the standard deviation of the frequency noise is 2\% of the FWHM. We find in the manual of the HP/Agilent 8360 series of sweepers (which is a widely used and representative microwave sweeper oscillator) that for "Sweep widths $>n \times 10$ MHz: Lesser of 1\% of sweep width or $n \times 1$ MHz + 0.1\% of sweep width". We conclude that these value agree well with the frequency inaccuracy which we deduced from the Monte Carlo simulations.

We note that the two types of noises, signal and frequency noise, are uncorrelated and can be simultaneously present, when their variance (square of standard deviations) are additive. However, we believe that typically one can reduce the signal noise to a low level and eventually the frequency noise dominates the observed error of the resonator parameters.

\section{Details of the Monta Carlo simulations}

We attempted to explain the observed noise in $f_0$ and $Q$ using a Monte Carlo method. We modeled the experimental situation as an exponential decay of a sinusoidal signal and used the same method to calculate the resonator parameters as from a transient provided by a measurement. The model of the transient:

\begin{gather}
V_I(t)=A\cdot\text{e}^{\frac{-2 \pi\cdot t \cdot f_0}{Q}}\cdot\text{cos}\left(2\pi\cdot f\cdot t+\phi\right)
\end{gather}
\begin{gather}
V_Q(t)=A\cdot\text{e}^{\frac{-2 \pi\cdot t\cdot f_0}{Q}}\cdot\text{cos}\left(2\pi\cdot f_0\cdot t+\pi/2+\phi\right)
\end{gather}

\noindent The following possible noise sources were modeled: 
\begin{itemize}
\item the instability of $Q$ observed between transients
\item the instability of $f_0$ observed between transients
\item the rapid frequency noise affecting $\phi$
\item voltage noise at the mixer output
\end{itemize}
We found the $Q$ and $f_0$ noises to give rise to very different measurement noises for the two measured parameters, while we found $\delta\left(Q\right)$ and $\delta\left(f_0\right)$ to be in the same order of magnitude. We found that a frequency noise of about 10 kHz implemented in $\phi$ gives rise to similar parameter noises as found in the measurements. However the frequency stability of the sources are higher than this value, so this is not the greatest source of measurement noise. The voltage noise predicted comparable parameter noises to their measured values, so we concluded that the measurement is dominated by voltage noise.

\section{Sensitivity of the resonator perturbation technique for the material parameters}

The improvement of the signal to noise ratio regarding the cavity parameters (eigen-frequency and quality factor) directly yields an improved precision of the determined material properties with the use of the use of cavity perturbation methods. Herein, we consider the electric permittivity but this can be straightforwardly applied for the magnetic permeability\cite{chen2004microwave}. The selection between the two types of measurables can be controlled whether the sample is placed in the node of the microwave magnetic field (electric field only) or the node of the microwave electric field (magnetic field only) \cite{PooleBook,chen2004microwave}, which can be conveniently achieved in e.g. a rectangular TE10n type cavity\cite{PooleBook}.

We previously established the proper measures of the error in the $Q$ and $f_0$ measurements as:

\begin{gather}
 \delta(f_0)=\frac{\sigma(f_0)}{ \Delta f}
 \end{gather}
\begin{gather}
 \delta(Q)=\frac{\sigma(Q)}{Q}
 \end{gather}
\noindent where $\Delta f$ is the FWHM of the cavity resonance and the $\sigma(.)$ denote the standard deviation of the respective quantity in the measurement. The merit of these error definitions is that these do not change with the $Q$ factor. 

The sensitivity of the cavity parameters for the material properties is an important factor determining the effectiveness of the whole measurement system. Herein, we present a figure of merit for material properties measured using a microwave resonator. The dielectric properties of a material can be obtained from the change in the eigen-frequency and quality factor of the resonator as \cite{chen2004microwave}:

 \begin{gather}
 \epsilon_\text{r}'-1=A \frac{V_\text{s}}{V_\text{c}} \frac{f_\text{e}-f_\text{s}}{f_\text{s}}
 \end{gather}
\begin{gather}
 \epsilon_\text{r}''=A \frac{V_\text{s}}{V_\text{c}} \left(\frac{1}{2Q_\text{s}}-\frac{1}{2Q_\text{e}}\right)
 \end{gather}
where $\epsilon_\text{r}=\epsilon_\text{r}'+\mathrm{i} \cdot \epsilon_\text{r}''$ is the complex dielectric constant of the studied material, $f_\text{e}$ and $Q_\text{e}$ are the eigen-frequency and quality factor measured without the sample (also known as empty or unloaded values) and $f_\text{s}$ and $Q_\text{s}$ values are those measured with sample. $V_\text{s}$ and $V_\text{c}$ denote the sample and cavity volumes. The constant $A$ is related to the measurement configuration and working mode of the cavity, the shape and location of the sample in the cavity. A good approximation for $A$ reads:

\begin{gather}
A=\frac{\int E_\text{e}^*\cdot E_\text{s}\, \text{d}V}{\int
\lvert{E_\text{e}} \rvert^2\text{d} V},
\end{gather}
where $E_\text{e}$ and $E_\text{s}$ denote the electric field in the microwave cavity in the absence and presence of the sample, respectively.

In case the presence of the sample little perturbs the cavity, $f_\text{s}\approx f_\text{e}$
\begin{gather}
\epsilon_\text{r}'-1\approx\frac{f_\text{e}-f_\text{s}}{f_\text{e}} \frac{V_\text{s}A}{V_\text{c}}
\end{gather}

The sample volume when expressed with the mass, $m_{\text{s}}$

\begin{gather}
V_\text{s}=\frac{m_\text{s}}{\rho_\text{s}}
\end{gather}

\noindent where $m_\text{s}$ and $\rho_\text{s}$ are the the mass and density of the sample.

A constant, which expresses the sensitivity of the measurement geometry, $K$ can be defined, which reflects the frequency shift weighted by sample mass, density and dielectric constant, as follows:
\begin{gather}
K=\frac{f_\text{e} A}{V_\text{c}}=\frac{\left(f_\text{e}-f_\text{s}\right) \rho_\text{s}}{\left(\epsilon_\text{r}-1\right) m_\text{s}}
\end{gather}

As mentioned earlier, the error of the frequency measurement, $\delta(f_0)$, is a value which is more or less the same for different microwave resonators, and is mainly determined by the approach chosen to measure $f_0$ and $Q$. It allows to give another factor that ultimately defines the setup sensitivity including the sensitivity of the geometry to the sample and the precision of the measurement for the resonator parameters:

\begin{gather}
 \frac{K}{\delta(f_0)}=\frac{\Delta f}{\sigma (f_0)} \frac{\left(f_\text{e}-f_\text{s}\right)}{\left(\epsilon_\text{r}'-1\right)} \frac{\rho_\text{s}}{m_\text{s}}
 \end{gather}
This gives an indication about the amount of the sample required for a given sensitivity in the material parameters. Either an increase in $K$ or a decrease in $\delta(f_0)$ improves the signal to noise ratio of the measurement or it allows to reduce the required sample amount in the study.

\section{Analogy between the Gabor uncertainty and the error in the $Q$ and $f_0$ measurement}

We would like to point out a compelling analogy between the error benchmark defined in Ref. \onlinecite{GyureRSI} and the so-called Gabor uncertainty\cite{Gabor1946,CohenTimeFreq,Roy1996,Poisson1999,Oliviera2000,Hall2006,Dodonov2015}. The latter states that for a a signal with a given time, $\sigma_t$, and frequency uncertainty, $\sigma_\omega$ it holds:

\begin{gather}
\sigma_t\times \sigma_\omega \geq 1/2.
\label{eq:Gabor_uncert}
\end{gather}
The equality in Eq. \eqref{eq:Gabor_uncert} holds only when the signal is a Gaussian in both time and frequency domain. We recognize that in the definition of $\delta\left(f_0\right)$, the $\sigma \left( f_0\right)$ numerator can be identified as $\sigma_\omega/2\pi$ and the bandwidth denominator, $\Delta f$ can be rewritten as uncertainty of the time measurement: $\Delta f=1/2\pi\tau=1/2\pi\sigma_t$. Therefore we formally obtain:

\begin{gather}
\delta(f_0)=\sigma_t\times \sigma_\omega \geq 1.
\label{eq:Gabor_uncert2}
\end{gather}

Formally, our result of $\delta f_0<10^{-4}$ appears to violate the Gabor uncertainty. This is however artificial as the Gabor uncertainty expresses the expected standard deviation of $\omega$ not the error with which it can be measured. We envisage a source with a perfect, noiseless oscillation at $f_0$. If it is measured in a 1 sec long Gaussian window, and the time domain signal is Fourier transformed, the result is a Gaussian with $1/2\pi$ Hz bandwidth. However, the \emph{position} of the Gaussian can be determined with a much larger accuracy than 1 Hz, the Gabor uncertainty expresses only that the center lies within this domain with a probability of 1. Another aspect is that the Gabor uncertainty is mainly valid for propagating (information carrying) signals, which is not the case herein.


\end{document}